\documentclass[11pt]{article}
\usepackage{amssymb}
\usepackage{graphicx}
\usepackage{amsmath}
\newcommand{\be}{\begin{equation}}
\newcommand{\ee}{\end{equation}}

\begin{document}
\baselineskip18pt
\title{Information functionals and the notion of  (un)certainty: RMT - inspired case}
\author{Piotr Garbaczewski\thanks{Presented at the 3rd Workshop on Quantum Chaos and Localization Phenomena,
Warsaw May 25-27,2007;  Email: pgar@uni.opole.pl} \\
Institute of  Physics,   University of Opole, 45-052 Opole,   Poland}
\maketitle
\begin{abstract}
Information functionals allow to  quantify the degree of randomness of a given probability distribution,
either absolutely (through min/max entropy principles) or relative to a prescribed
 reference  one.  Our primary aim  is to  analyze  the "minimum information" assumption, which is
  a classic concept (R. Balian, 1968)  in the random matrix
  theory. We put  special  emphasis  on generic level (eigenvalue) spacing  distributions and the degree of their
  randomness,  or alternatively - information/organization deficit.
 \end{abstract}

 \noindent


\section{Motivation}

The statistical theory of random-matrix spectra \cite{haake,mehta} provides an ideal
 playground to test workings of the   Shannon and Kullback -Leibler  entropies in diverse  contexts.
 That pertains to  a direct  analysis of  spectral  data for complex quantum systems (semiclasically chaotic case
 included), but as well to the statistics of   Gaussian  matrix ensembles and  random matrix  diffusion processes.
   Dyson's interacting Brownian motion model can be interpreted as  as a non-equilibrium dynamical  process,
  whose asymptotic   distribution is  related to   the  thermodynamical   equilibrium  state of  a  Coulomb  gas
  (RMT as equilibrium statistical mechanics).
    Ultimately  one  may pass to  probability densities inferred from the ground state(s)  of   singular  Calogero-type
quantum systems:  Shannon and K-L entropies prove to be proper tools
in the quantum case as well.

Before embarking on these issues, let us  indicate  that there are
ambiguities  involved in the very concept of \it   information \rm
    and \it  (un)certainty\rm.    To stay on a  solid  ground, \cite{cover}-\cite{gar}, we must
accept a specific lore of semantic games, where baffling synonyms
quite often appear and their specific  meaning is under  scrutiny.
Examples are: information vs  entropy notions, (un)certainty  and
randomness vs information deficit, entropic measures of surprise vs
information functionals,   min/max  entropy principle vs  effective
randomness (uncertainty),   uncertainty  (lack of information) vs
(quantum)  indeterminacy.

Since a particular definition of an  entropy  functional  is
non-unique and to  high extent purpose-dependent \cite{gar} one must make
suitable  choices in the entropic  menu ("entropic mess", with a
partially random order of entries):  Clausius thermodynamic entropy,
Boltzmann, Gibbs,  Shannon, relative, conditional, Kullback-Leibler,
Renyi, Tsallis, Wehrl, information entropy, differential entropy,
Kolmogorov-Sinai entropy, von Neumann entropy; the list may be
continued.

We shall basically invoke the Shannon and Kullback-Leibler
entropies, in conjunction with continuous probability distributions
on $R^+$. We   base  our  discussion on the text-book  wisdom  that
the  entropy  is  a measure of the degree of randomness  and the
tendency  (in the time domain) of physical systems to become less
and less organized. We extend this verbal phrase to
probability densities of the   functional  form:
\begin{equation}
 f(x) \sim s^{\beta } \exp(-{s^{\alpha }})
\end{equation}
 with  $s \in R^+$,   $\alpha = 1$ or $2$,  while  $\beta =0, 1, 2, 3, 4$.

The above formula encompasses \cite{haake} a number of  "quantum chaos"-related
level spacing distributions: Poisson (strictly speaking
-exponential), semi-Poissonian   of various types and the generic family
of spacing densities, that are  exact for   $2\times 2$  random
matrices, and are identifiable as $n=2,3,4,5$
Bessel-Ornstein-Uhlenbeck probability  laws (densities)  on $R^+$.

The latter arise directly from Gaussian matrix ensembles   and $n$ in  the exponent of ${ s}^{n-1}$ counts the  independent
$2\times 2$  Gaussian random matrix elements, ($\beta = n-1$).
 With   the  $\langle s\rangle =1$ normalization, we have:

$$P_{GOE}(s) =   { s} {{\pi }\over 2} \exp(-s^2{\pi \over 4})$$

\begin{equation}
P_{GUE}(s) = { s^2} {{32}\over \pi ^2} \exp(-s^2 {4 \over \pi })
\end{equation}

$$P_{Ginibre}(s) = { s^3} {{3^4\pi ^2}\over  {2^7}}  \exp(-s^2 {{3^2\pi }\over {2^4}})$$

 $$P_{GSE}(s) = { s^4} {{2^{18}}\over {3^6 \pi ^3}}\, \exp(- s^2 {64 \over {9\pi }})$$

The $\beta =0$, $\langle s\rangle =1$  normalized  Gaussian on $R^+$
reads $P_0(s)=  (2/\pi )\exp (-s^2 / \pi )$, and has  variance  $\langle (s -\langle s\rangle )^2\rangle = (\pi - 2)/2$.

\section{Random variables  on $R^+$ and entropic measures of probability (de)localization}

\subsection{Entropies}
Given a probability measure
 $\sum_{j=1}^N \mu  _j = 1$.   Its Shannon entropy reads   ${\cal{S}}(\mu ) = - \sum_{j=1}^N  \mu _j \ln \mu _j$
 and  takes  a maximum  value  $\ln\,  N$  in the "most random" case of  a   uniform  distribution:   $\mu _j = 1/N $ for all $1\leq j  \leq
 N$.
 An obvious minimum at  $0$  appears if for  any $j$   we have $\mu _j =1$.

We shall focus on continuous probability distributions on $R^+$. The corresponding  Shannon entropy is introduced as follows:
\begin{equation}
\int \rho (s) \, ds = 1  \rightarrow
{\cal{S}}(\rho)=   - \int \rho (s) \ln \rho (s)  dx
\end{equation}

At this point it is instructive to mention that in the realistic (spectral data analysis) "quantum chaos" framework, one encounters spacing histograms and
definitely \it not \rm continuous probability densities. The latter may merely be interpreted as useful continuous approximants  of
discrete probability measures.

The situation is more involved in case of the corresponding Shannon entropies, where the approximation issue is delicate.
Even if one  follows a pedestrian reasoning, we can justify and keep under control  the limiting behavior, \cite{cover,gar}:
\begin{equation}
  \sum_1^N \mu _j =1 \rightarrow \int \rho dx = 1\, .
\end{equation}
An immediate question is: what can be said about the  mutual  relationship of $ S(\mu ) =  - \sum_1^N  \mu _j  \ln \mu _j $
and $S(\rho)=   - \int \rho (s) \ln \rho (s) ds $ ?

We first observe that  $0 \leq - \sum_1^N \mu _j \ln \mu _j \leq \ln N$ and consider  an interval of length  $L$ on a line  with the a priori chosen
 partition   unit  $\Delta s = L/N$.
  Next, we define: $\mu _j \doteq  p_j \Delta s $  and notice that (formally, we bypass an issue of dimensional quantities)
\begin{equation}
S(\mu ) = - \sum_j  (\Delta s)  p_j \ln p_j  - \ln (\Delta s)
\end{equation}

Let us fix $L$ and allow   $N$ to grow, so that   $\Delta s $  decreases and the partition becomes finer. Then
\begin{equation}
 \ln (\Delta s) \, \,  \leq  \, \,   - \sum_j  (\Delta s)  p_j \ln p_j \, \,
 \leq \, \,  \ln L
\end{equation}
where
\begin{equation}
S(\mu ) + \ln (\Delta s)  =    - \sum_j  (\Delta s)  p_j \ln p_j  \Rightarrow
S(\rho)=   - \int \rho (s) \ln \rho (s) ds
\end{equation}
\vskip0.5cm
$S(\rho )$ is the Shannon  information  entropy  for the probability measure on the
interval $L$.  In the infinite volume $L\rightarrow \infty $ and
infinitesimal grating $\Delta s \rightarrow 0$ limits, the density functional $S(\rho )$  may be
unbounded both from below and above, even non-existent, and seems to  have lost any computationally  useful link with its
 coarse-grained version $S(\mu )$.

 However, the situation is not that bad, if we invoke standard methods \cite{cover,gar} to overcome a dimensional difficulty,
 inherent in the very definition of
$S(\rho )$, if we admit dimensional units. Namely, we can from the start take   a (sufficiently small) partition unit
 $\Delta s$ to have dimensions of length. We allow $s$ to carry length  dimension as well. Then, the dimensionless expression
 for the Shannon entropy of a continuous probability distribution  reads:
 \begin{equation}
S_{\Delta }(\rho)=   - \int \rho (s) \ln [ \Delta s \cdot  \rho (s)] ds
\end{equation}
and all of a sudden, a comparison of Eqs.(5) and  (8) appears to make sense. We can legitimately set  estimates for
$|S(\mu ) - S_{\Delta }(\rho )|$ and  directly verify the approximation validity  of  $S(\mu )$ in terms of $S_{\Delta }(\rho )$,
when the partition becomes finer.

In the present paper we are interested in properties of  various  continuous probability distributions, and \it not \rm their coarse-grained
versions. Therefore our further discussion will be devoid of any dimensional or partition unit connotations.  Since negative values
of the Shannon entropy are  now  admitted, instead of calling it an information measure, we prefer to tell about a  "localization measure",
 "measure of surprise" or "measure of information deficit".

\subsection{ Poissonian  spacing  distributions}
\vskip0.3cm
Let $X_1, X_2,...$ be independent random
variables  on $R^+$,  with a  common for all of them exponential probability law
\begin{equation}
\mu (x)= \alpha \exp(-\alpha x)
\end{equation}
$\alpha >0$ ,  mean $1\over \alpha $, variance
${1\over {\alpha ^2}}$.
 Let us denote $S_n = X_1 + X_2 +...+X_n$, $n=1,2,...$  and note that
 $S_n$  has the   density (Poisson probability  law):
\begin{equation}
p_{n}(x) = {{\alpha ^n\, x^{n-1}}\over {(n-1)!}}\exp(-\alpha x) \label{erlang}
\end{equation}
coming  from an  (n-1)-fold convolution of exponential probability
densities on $R^+$. The  law  is infinitely  divisible:
\begin{equation}
p_{n+m}(x) = (p_n*p_m)(x) = \int_0^x p_n(x-y) p_m(y) dy
\end{equation}
with $p_1(x)=\mu (x)$  and $n,m = 1,2,...$.

In particular,  $X_i+ X_j$   for any $i,j, \in N$
has  a probability density $p_2(x)={\alpha ^2}\, x\,  \exp(-\alpha x)$
which upon setting $\alpha =2$ and $x=s$  stands for an example of
a semi-Poisson law
\begin{equation}
P(s)=4 { s}\, \exp(-2s)
\end{equation}
known  to govern the  adjacent level
statistics for  a subclass of pseudo-integrable systems.
Other (plasma-model related) { semi-Poisson laws}
arise as well.
For example,  $S_3$ has a density  $p_3(x)$ which upon setting $\alpha =3$ and $x=s$,
  gives rise to
\begin{equation}
  P(s)={{27}\over 2} { s^2} \, exp(-3s)\, .
\end{equation}
Analogously, $S_5$ yields  $p_5(x)$ and upon setting $\alpha =5$
implies
\begin{equation}
P(s)= {{3125}\over {24}} { s^4}\, \exp(-5s)
\end{equation}

The distribution Eq.~(\ref{erlang}), here identified as the Poisson probability law for the random variable $S_n$,  in the information-theoretic
literature  is known as the  $(\alpha ,n)$-Erlang distribution. Its Shannon entropy reads, \cite{cover}:
\begin{equation}
{\cal{S}}(p_n)=   \ln \Gamma (n) + (1-n) \psi (n)   + n - \ln \alpha
\end{equation}
where  the Euler gamma function  $\Gamma (x) = \int_0^{\infty } \exp(-t)\,  t^{x-1} dt$ appears, together with
the digamma function (logarithmic derivative of $\Gamma $)  $ \psi (x) = {\frac{d}{dx}} \ln  \Gamma (x)$.

 We have  $\Gamma (n) = (n-1)!$ and $\psi (n) =
H_{n-1} - \gamma$, where $\gamma = \lim _{n\rightarrow \infty } (H_n - \ln n) \sim  0, 577215$ is the Euler-Mascheroni constant,
 while harmonic numbers $H_n=\sum_{k=1}^n (1/k) $
take the consecutive  values  $1, 3/2, 11/6, 25/12$ etc.

Notice that  $\alpha = n$ should
be set if one needs  to address the previous  $P(s)$. For the pure exponential law, we have: ${\cal{S}}(p_1)= 1 - \ln \alpha $ and the fit $\alpha =1$
would give us ${\cal{S}}(p_1)=0$.

\subsection{Bessel-Ornstein-Uhlenbeck processes and their  invariant densities}

Let $X_1, X_2, ...,X_n$ be   independent random variables with  common for all,  zero mean  and variance $1$,
Gauss (Brownian) probability law on $R$:
\begin{equation}
p(x) =  {\frac{1}{\sqrt{2\pi }}} \exp (- x^2/2)
\end{equation}
Let us consider
\begin{equation}
R_n\doteq  (X^2_1 + ... + X^2_n)^{1/2}
\end{equation}

Assume the Brownian motion (Wiener process) to proceed,  in $n$  independent copies. The  radial Brownian motion ({ Bessel  process})
is thereby  induced  on $R^+$.
The  probability  density of $ R\doteq R_n$, $n>1$    at time $t\in R^+$ is denoted by
 $\rho (r,t)$, $r\in R^+$. We have:
\begin{equation}
dR = ({{n-1 }\over {2R}})dt + dW  \,   \Longrightarrow   \,
\partial _t\rho  = {1\over 2} \triangle  \rho   - \nabla [{{(n-1) }
\over {2r}} \rho ]
\end{equation}
It is known that the point $r=0$  is never reached with the probability $1$, which models a { repulsion}, \cite{gar1}.  (Here,
$r=0$ is the so-called entrance boundary.)

If we impose  a {  restoring}  harmonic   force (proportional to a  randomly taken value of the  distance  $R_n$  from the origin).
\begin{equation}
dR = ({{n-1 }\over {2R}} - R)dt + dW \,
 \Longrightarrow \,
\partial _t\rho  = {1\over 2} \triangle  \rho   - \nabla [({{n- 1}
\over {2r}} - r)\rho ]
\end{equation}

We take  $\rho _0(r)$ with $r\in R^+$ as  the density of distribution
of the random variable  $R$  at time $t=0$. Then  the function $\rho (r,t)$, solving  the F-P equation,
is the density of $R= R(t)$ for all  $t>0$.

The $n>1$ family of  time
homogeneous  radial  (Bessel)  Ornstein-Uhlenbeck processes  is driven by
{ transition probability densities}, \cite{karlin}:
\begin{equation}
p_t(r',r) = p(r',0,r,t) =
2 r^{n-1} \exp(-r^2)\cdot
\end{equation}
$$
{1\over {1-\exp(-2t)}}  \exp[-{{(r^2 + {r'}^2)\exp(-2t)}\over {1-\exp(-2t)}}]\cdot
[r r' \exp(-t)]^{-\alpha }
I_{\alpha }({{2r r'\exp(-t)}\over {1-\exp(-2t)}})
$$
where $\alpha = {{n-2}\over 2}$ and $I_{\alpha }(z)$ is a modified Bessel function
of order $\alpha $:
\begin{equation}
I_{\alpha } (z) = \sum_{k=0}^{\infty } {{(z/2)^{2k+\alpha }}
\over { (k!) \Gamma (k+\alpha + 1)}}
\end{equation}
We recall special values of the Euler gamma function:  $\Gamma (n+1) = n!$ and
$\Gamma (n + 1/2) =   (2n)!  \sqrt {\pi }/ n! 2^{2n}$.

Straightforwardly,  one can verify that asymptotic  densities of the Bessel-OU  process have the  form:
\begin{equation}
 \rho _*(r) =  {2\over {\Gamma (n/2)}}{ r}^{n-1} \exp(-r^2)  \label{BOU}
\end{equation}
A complementary  check amounts to observing that  the forward drift $b(r)$  of the  stationary  B-OU  process needs to  obey
  $\partial _t \rho _* = (1/2) \Delta \rho _* - \nabla (b\, \rho _*)=0$. The invariant (asymptotic) density   reads:
\begin{equation}
\rho _*(r) = {\frac{1}{Z}}  \exp (-V)
\end{equation}
 with the normalization $Z= \int_{R^+}  \exp (-V) dr $. We have
 \begin{equation}
V  = V(r)=   {\frac{1}{2}} [r^2 -(n-1) \ln r ]
\end{equation}
and
\begin{equation}
b(r) =   {\frac{1}{2}} \nabla \ln \rho _*(r) =  - \nabla V = {\frac{n-1}{2r}} - r \, .
\end{equation}
After normalizing the mean, $\langle R \rangle  = 1$, and replacing  $r$ by $ s$ we readily arrive at
the previous  RMT  spacing formulas.

The Shannon entropy  of the  continuous probability distribution ( B-OU family) Eq.~(\ref{BOU}) reads, \cite{cover}:
\begin{equation}
{\cal{S}}(\rho _*)= \ln \Gamma \left({\frac{n}2}\right)   - {\frac{n-1}2}\, \psi \left({\frac{n}2}\right)  + {\frac{n-1}2}
\end{equation}
where for half-integer values, the digamma function $\psi $  equals:
\begin{equation}
\psi (n+ {\frac{1}2})  = - \gamma - 2\ln  2  + \sum_{k=1}^n {\frac{2}{2k-1}}\, .
\end{equation}
For the  Gaussian on $R^+$, i. e. $ \rho _*(r)= (2/\sqrt{\pi }) \exp (- r^2)$, we have ${\cal{S}}(\rho _*)= (1/2) \ln \pi $.
It is useful to reproduce the general Shannon  entropy formula for the Gaussian on $R^+$:
\begin{equation}
\rho (r) = [2/ \pi \sigma ^2 ]^{1/2}  \exp [ - r^2/ 2\sigma ^2)] \Longrightarrow
{\cal{S}}(\rho ) = (1/2)[ \ln (\sigma ^2 \pi /2)  + 1]\, .
\end{equation}

\subsection{Calogero model}

For general  stationary diffusion  processes, a formula relating forward drifts $b(x)$
of the stochastic process
with potentials  ${\cal{V}}(x)$ of an auxiliary  conservative Hamiltonian system reads, \cite{gar,gar1} (we choose a diffusion
coefficient to be equal $1\over 2$, hence scale away $\hbar $ and $m$):
\begin{equation}
{\cal{V}}(x) = {1\over 2} (b^2 + \nabla \cdot b)\, .
\end{equation}
Upon substituting
\begin{equation}
b(x) =  {\frac{\beta }{2x}}  - x
\end{equation}
 with $\beta = n-1$  we arrive at:
\begin{equation}
{\cal{V}}(x) = {1\over 2} [ {{\beta (\beta -2)}\over {4 x^2}}  + x^2]   - {1\over 2}(\beta +1)
\end{equation}
 This potential function enters a standard definition of the one
particle Hamiltonian operator (no physical parameters):
\begin{equation}
H = -{1\over 2} \triangle   + {\cal{V}}(x)
\end{equation}
where $\triangle = {{d^2}\over {dx^2}}$.  The  energy operator $H$,   with  the previously introduced  ${\cal{V}}(x)$,
is an equivalent form of  a two-particle  (actually two-interacting-levels) version of the
Calogero-Moser Hamiltonian, \cite{haake,gar1,calogero}.

The classic Calogero-type problem is defined by
\begin{equation}
H = - {1\over 2}{d^2\over {dx^2}} + {1\over 2}x^2 + {{\beta (\beta - 2)} \over {8x^2}}
\end{equation}
with the well known spectral solution:
\begin{equation}
E_k(\beta )
= 2k + 1 + {1\over 2}[1+\beta (\beta - 2)]^{1/2}
\end{equation}
where  $k\geq 0$ and $\beta > -1$.
By substituting $\beta = 1,2,3,4$ we  easily check that  $E_0(\beta ) = {1\over 2}(\beta + 1)$.

All previously considered $n= 2,3,4,5$ radial diffusion  processes
correspond to Calogero-Moser   potentials and thence   Calogero  operators in  the (renormalized) form  $H - E_0$
 where $E_0$ is the  respective (fix $n$)  ground state  (k=0) eigenvalue.
These stochastic processes   arise as  the so-called  ground
state processes  associated with the  Calogero Hamiltonians.  (Note: we are aware of all the "fictitious time" Dyson's model
philosophy):  if $ \psi _0$ is the ground state { wave function}, we regard  $ \rho _* \doteq |\psi _0|^2$ as
 an invariant  { probability density} of the stochastic B-OU process.
 Let us recall  that
the classic Ornstein - Uhlenbeck process can be regarded as  the ground
state process of the harmonic oscillator Hamiltonian operator.

\subsection{General comments on the  quantum  Calogero  system}
The Calogero {  singular} quantum mechanical  Hamiltonian
\begin{equation}
H = - {d^2\over {dx^2}} + x^2 + {\gamma \over {x^2}}
\end{equation}
has the  eigenvalues
 $$E_n = 4n + 2 + (1+4\gamma )^{1/2}$$
 where  $n\geq 0$ and $\gamma >
-{1\over 4}$.
The  eigenfunctions have the form:
$$
f_n(x) = x^{(2\alpha +1)/2} \exp(-{x^2\over 2})\,
L^{\alpha }_n(x^2)
$$
with $\alpha = {1\over 2}(1+4\gamma )^{1/2}$ and
\begin{equation}
L_n^{\alpha } (x^2) = \sum_{\nu =0}^n {{(n+\alpha )!}\over {(n-\nu )!
(\alpha + \nu )!}} {{(-x^2)^{\nu }}\over {\nu !}}\, .
\end{equation}
The $\gamma $ parameter range $-1/4< \gamma < 3/4$ involves some
mathematical subtleties  concerning  the singularity at $0$, which is {  not}
sufficiently severe to  enforce the Dirichlet boundary
condition.

In the range  $\gamma \geq 3/4$ we deal with  a
{ double degeneracy} of the ground state and of the eigenspace of the
self-adjoint operator $H$.
The singularity at $x=0$  { completely  decouples} $(-\infty
,0)$ from $(0,+\infty )$ so that $L^2(-\infty,0)$ and $L^2(0,+\infty )$
are  the invariant subspaces for   the unitary  Schr\"{o}dinger evolution
 $\exp(-iHt)$  generated by $H$.
The related Schr\"{o}dinger probability
current  vanishes  at $x=0$ for all times and { there is no dynamically
implemented communication} between those two areas, c.f. \cite{karw}.
 The respective localization probabilities,  to find a particle on a  positive or negative semi-axis,  are constants
of motion. Because of the singularity
at $0$, { once trapped}, a particle is confined  in one    particular
enclosure only and then cannot be detected in another.

 The (positive semi-axis)  projection operator   $P_+$ defined by $(P_+ f)(x)=
\chi_{R^+}(x) f(x)$  commutes  with $H$. It is thus  tempting  and (with suitable precautions)
 legitimate  to confine the
discussion to $R^+$  (or $R^-$) {  separately}.
However, we can not   tell here
about two {  disjoint } quantum problems defined  respectively  on
$R^+$ and $R^-$. We deal  with a { single} quantum mechanical system,
though technically - with a { degenerate ground state}.

Let us also  point out that $D(H)$ contains functions restricted to
obey  $f(0)=0=f'(0)$ and not necessarily to vanish on any of half-lines.
Such functions
 { may have  support on both}, positive and negative semi-axes
simultaneously, excluding
  the  origin $0$.
For example, a normalized linear combination  (standard {  superposition})  of the  two components of the
 degenerate ground state of $H$, is a legitimate element of
$D(H)$.   There is no mixture in here.
Are they very special  Schr\"{o}dinger cat states ? -  good lurking-place for  the cat-metaphysics ?

\subsection{  Dyson's asymptotic equilibrium}
\vskip0.5cm
\noindent
Let  $M$ be a  Hermitian  $n\times n$ matrix with an orthogonal, unitary or  symplectic invariance built-in. Then, the
 number of independent matrix elements equals, respectively $N=n+ {\frac{1}{2}}n(n-1) \beta $,
 $\beta = 1, 2, 4 $.
 We introduce a
Gaussian matrix  ensemble: independent matrix elements are interpreted as { independent} Gaussian random variables
with zero mean and variance, \cite{dyson}:
\begin{equation}
Var (M_{ij}) =
{\frac{a^2}{2\beta }} (1+ \delta _{ij})\, .
\end{equation}
 The probability density reads
\begin{equation}
P_*(M_1,...., M_N) = c \, \cdot \exp [- \beta\,  Tr(M M^*)/2a^2]\, .
\end{equation}
The  Gaussian  RM  joint  density  of (real)  eigenvalues   has the form
\begin{equation}
\Lambda _*(x,_1,x_2,...x_n) = C \cdot  [\prod _{i<j} |x_i - x_j |^{\beta } ]  \exp[- \beta (\sum_i x^2_i)/2a^2]=
\end{equation}
$$
C \exp [-\beta   (- \sum_{i<j} \ln |x_i - x_j|  + \sum_i {\frac{x^2_i}{2a^2}})]
$$
i. e. (c.f. Eqs. (23) and  (24))
$$
\Lambda _* = {\frac{1}{Z}}  \exp [-\beta V]
$$
\begin{equation}
V=V(x_1,x_2,...,x_n) = - \sum_{i<j} \ln |x_i - x_j|  + \sum_i {\frac{x^2_i}{2a^2}}
\end{equation}
The above observations can be inferred by passing to  suitable  reduced densities, in the   asymptotic limit
 of the   Smoluchowski  diffusion equation  for the time-dependent
probability density $P(M_1,M_2,...,M_N, t)$:\\
\begin{equation}
\partial _t P =  \sum_{M= M_{ij}}  \left[ {\frac{(1+ \delta _{ij})}{\beta \nu }} {\frac{\partial ^2}{\partial M^2}} P
+ {\frac{1}{a^2}} {\frac{\partial }{\partial M}} (M\, \cdot P)\right]
\end{equation}

Let us fix an initial $t=0$ condition to be $M'$,  $\nu $ is an auxiliary  "friction"
coefficient. We deal with independent Ornstein-Uhlenbeck processes  for random  matrix elements
\begin{equation}
P(M,t)  = c\cdot [1- q^2]^{-N/2} \cdot \exp  \left[ -   \beta {\frac{ Tr(M- qM')^2}{2a^2(1-q^2)}}\right] \rightarrow P_*(M)
\end{equation}
$$q=\exp[-t/ a^2 \nu ] $$
which, in turn  turn, induces the corresponding  interacting Brownian motions for the eigenvalues. $P_*$ and  $\Lambda _*$
stand for  invariant asymptotic (unique stationary, equilibrium) densities of these processes.

Notice that choosing $a^2= \beta n$, we can write down a corresponding set of stochastic differential equations
(infinitesimal increments) for the interacting Brownian motions associated with teh $n\times n$ random matrix. Forward
drifts  read  $b_j(\lambda ,t) = =  - \nabla _j  (\beta V)$ and thence:
\begin{equation}
d\lambda _j(t) = \left[ - {\frac{1}{2n}} \lambda _j  + {\frac{\beta }{2}} \sum_{i<j} {\frac{1}{\lambda _j(t) - \lambda _i(t)}}\right] dt + dW_j(t)
\end{equation}
Their properties were studied in detail in the mathematical literature and despite the Coulomb (or centrifugal) singularity, the solution is
known to be unique and non-explosive for all times and all $n$, including the  $n\rightarrow \infty $ limit. The eigenvalues
 never cross and an initially given  (time  $t_0$)   order $\lambda _1 < \lambda _2 < ...<\lambda _n$ of  (real) eigenvalues   is kept
  forever in the course of the diffusion process, \cite{rogers}.

  For clarity of discussion, let us illustrate a passage from $P_*$ to  spacing distributions $P(s)$ for $2\times 2$ random matrices.  Namely,
given the $n=2$ form of Eq. (24)
\begin{equation}
P(M_1,...,M_N) = c \, \cdot \exp \left[- {\frac{\beta}4}  (E_1^2 + E_2^2) \right]
\end{equation}
and
\begin{equation}
\Lambda _*(\lambda _1, \lambda _2) = C\cdot  |\lambda _1 - \lambda _2| \cdot \exp \left[- {\frac{\beta}4}  (\lambda _1^2 +  \lambda _2^2) \right]
\end{equation}
yield the  spacing distribution as a reduced density:
\begin{equation}
P(s) = const  \int d\lambda _1 \int d\lambda _2 \,  \delta (s- |\lambda _2-  \lambda _1|) \cdot \Lambda _*(\lambda _1, \lambda _2)\, .
\end{equation}
which upon securing  the normalization of $P(s) $ on $R^+$ and    $\langle s\rangle = 1$, gives rise to the generic RMT-spacing densities.

\section{Extremum principles: what can we say about the  degree of randomization ?}

\subsection{ Min/max    information entropy principle in the RMT}
The information-theory route, according to R. Balian, \cite{balian,mehta}, begins  as follow. We consider a constrained extremum   for a
 functional of a convex  function:
\begin{equation}
I = \int d\mu  (M) P(M)\cdot  ln \, P(M)
\end{equation}
with constraints $ \int d\mu (M) P(M) =1$  and
\begin{equation}
 \langle  Tr(M M^*) \rangle =    \int d\mu (M)\,  Tr(M M^*) P(M) = N  a^2
 \end{equation}
Next, one passes to  an extended information functional with Lagrange multipliers $b$, $d$
\begin{equation}
I(b,d) = \int d\mu (M) [  P(M) ln \, P(M) + b P(M) + d \cdot Tr(MM^*) P(M)]
\end{equation}
\vskip0.5cm
and looks for an {  extremum}  of $I$ (actually for  a   {   minimum information measure}, c.f. convexity property),   under the imposed  constraints.
The outcome is:
\begin{equation}
P_*(M) = \exp [-(1+b+d\cdot  Tr(MM^*))]
\end{equation}
with $ d = {\frac{1}{2a^2}} $  and $\exp[-(1+a)] =  c =  \left({\frac{1}{2\pi a^2}}\right)^{ N /2}$. We have  thus
 arrived at the  invariant  (Gaussian) probability measure  $P_*(M)$ for  the  Gaussian matrix ensemble, c.f. also \cite{haake,yukawa,hasegawa,muttalib}.

\subsection{ Maximum randomness issue: constrained  extremum}

To have a better insight into the  extremum principles at work, let  us recall the standard
maximum (information/Shannon)  entropy principle:  consider $[a,b] \in R$, assume that everything   you know  about the  a priori
 unknown probability measure   are (possibly) its moments
\begin{equation}
\int_a^b x^k \rho (x) dx = m_k
\end{equation}
with $ k=0, 1,..., M$ and  $m_0=1$-the normalization condition.

We look for densities that maximize the Shannon entropy of a continuous probability distribution (now
we encounter a functional of a concave function):
\begin{equation}
{\cal{S}}[\rho ] =- \int_a^b \rho \ln \rho dx
\end{equation}
under the constraint  of  $M$ fixed moments, \cite{sobczyk,mead}.

The extremum of a functional
\begin{equation}
\tilde{{\cal{S}}}  =- \int_a^b \rho \ln \rho dx  + \sum_0^M \lambda _k (\int_a^b x^k \rho dx - m_k)
\end{equation}
(a concavity property  of ${\cal{S}}$ needs to be rememebered) sets  the functional form of $\rho $  which  maximizes  the entropy:
\begin{equation}
\rho _*(x)= C \exp [- \sum_a^b \lambda _k  x^k]
\end{equation}
where  $C= \exp(-\lambda _0 -1)$ is the normalization constant  and  $\lambda _k$'s  are  fixed by identities
\begin{equation}
\int_a^b x^k \exp [- \sum_a^b \lambda ^k  x^k]\, dx= m_k
\end{equation}
If there is a { unique } solution in terms of  $\lambda _1,...,\lambda _M$, we say that
that an entropy  maximizing (under the $m_k$ "circumstances") density does exist.

For reference, let us reproduce some pieces of a standard wisdom:\\
(i) If $a$ and $b$ are finite, there exists a unique  maximum  entropy density;\\
(ii) In   $ R^+$ e.g. $ [0, +\infty )$ a maximizing density exists if   $m^2_1 \leq m_2 \leq 2m_1^2$.\\
{\it Notes:} if there is no constraint there is no maximizing density;  if only the mean $m_1=  1/\alpha   $ is given, we get
 the { exponential} one: $\rho _*(x)=\alpha \exp (- \alpha x)$; for the   Gaussian on $R^+$, like e.g.
 $ \rho (r)= (2/\sqrt{\pi }) \exp (- r^2)$   ,
 we have ${\cal{S}}(\rho )= (1/2) (\ln \pi  + 1)$ which is a maximum of the Shannon entropy under the moment  constraints
 $m_1= \langle r \rangle =1/ \sqrt{\pi }$ and $m_2=\langle r^2\rangle = 1/2$; for another   Gaussian on $R^+$,
 $P_0(s)=  (2/\pi )\exp (-s^2 / \pi )$, we have  $m_1 =\langle s\rangle =1$,
   $m_2=\langle (s -\langle s\rangle )^2\rangle = (\pi - 2)/2$ and  ${\cal{S}}(\rho )= (1/2)[ \ln (\pi ^2/4)   + 1]$.\\
(iii)  In $R$,  with no moment prescribed,  or given  the mean only,   there is no maximum entropy density.\\
{\it Notes:} if $m_1$ and $m_2$ are given, the maximum entropy distribution is the normal (Gaussian) one, with variance
$\sigma ^2 = m_2 - m_1^2$ i. e. $\rho (x) =  {\frac{1}{\sqrt{2\pi }\sigma } } \exp [- (x- m_1)^2/2\sigma ^2]$  and the Shannon entropy
value is ${\cal{S}}(\rho )= (1/2) \ln (2\pi e \sigma ^2)$.  That is to be compared with the previous outcome, Eq.~(28),  for the Gaussian on $R^+$.

\subsection{Extremum principles: statement of purpose}

{ Our RMT/Bessel-OU}  (spacing) distributions   fall into the above category { (ii)} where
 an absolute (least constrained)  entropy maximum on $R^+$  is  set by the exponential density.
With quite  a variety of probability densities on $R^+$ in hands, can we quantify their "randomness level"  \cite{gar3} absolutely,
  or relative to any of reference densities ?

{ Possible indications  towards  this end read as follows:}\\
(*)  compare absolute values of the   respective Shannon entropies for densities on $R^+$ (Cover-Thomas differential
 entropy tables);\\
(**) invoke Kullback-Leibler relative entropies and look for a minimum of the { relative entropy};
 relative with respect to the chosen  reference density;\\
 (***)  look for min/max principles   (like e.g. Helmholtz free energy properties)
 that govern the  time evolution of  standard diffusion-type processes (it is immaterial whether the time label
 is real or "fictitious");\\
 (****)  investigate the  "organization level" of     ground state densities of relevant quantum systems
  in comparison to  all possible eigenstate-related   probability densities. What about their "randomization" behavior
   with the growth of energy eigenvalues ?

In fact, before   we have established direct  comparison  tools,  e.g. Shannon entropy values  for the Poissonian and  generalized Gaussian
(B-OU, with a polynomial repulsion  factor) densities, c.f. Eqs.(15), (26) and (28). Hence the above point (*) has received due attention.

\subsection{  Kullback relative entropy route}

Concerning the point (**), let us define the relative entropy for densities on $R^+$:
\begin{equation}
I(\rho : \rho _{ref}) = \int \rho  \ln {\frac{\rho }{\rho _{ref}}}  dx
\end{equation}
with the  a priori prescribed $\rho _{ref}$.

Let us investigate \cite{kullback}  a  { minimum } of the
Kullback entropy $I(\rho : \rho _{ref})$ under the constraint imposed with the aid of an { auxiliary}  function $T(x)$:
\begin{equation}
\int T(x)  \rho (x) dx = \theta = const
\end{equation}
A  conditional  extremum  (minimum, in view of  the convexity property)   of a  functional
\begin{equation}
\tilde{I}  = \int [\rho  \ln {\frac{\rho }{\rho _{ref}}}   + \lambda  T(x)\rho + \lambda _0\rho ] dx
\end{equation}
is reached  at $\rho \rightarrow \rho _*$, with the constraint $\int T(x)  \rho _*(x) dx = \theta = const$:
\begin{equation}
\rho _*(x) =  C \,  \rho _{ref}(x) \,  \exp [- \lambda  T(x)]
\end{equation}

We need  an overall normalization  of $\rho _*$ on $R^+$. It is thus useful to   demand  that
we have fixed   an integral:
\begin{equation}
 {\frac{1}{C(\lambda )}}  = \int \rho _{ref}(x)\,   \exp [- \lambda   T(x)]dx
\end{equation}

{ Our guess:} choose $\rho _{ref}(x)$  and   $T(x)$, next adjust the values of $\theta$  and  $C(\lambda )$ to  fit
 either of  $\lambda  = 1, 2, 3...$.\\
Explicit examples are:\\
(i) Exponential family  ({ semi-Poisson} case)on $R^+$:\\
\begin{equation}
\rho _{ref} = \alpha \exp (-\alpha x)
\end{equation}
$$T(x) = - \ln x$$
(ii) Gaussian family  ({ Wigner} surmise) on $R^+$:\\
\begin{equation}
\rho _{ref} = (1/\sqrt{\pi }) \exp (-x^2)
\end{equation}
$$T(x) = - \ln x$$

The only delicate computational issue is the constraint Eq.~(57), where  logarithmic  integrations are to be carried out.
The  pertinent (reference)  integrals read:
\begin{equation}
\int_0^{\infty }   \exp(- \alpha  x) \,  \ln x \,  dx =  -{\frac{1}\alpha } (\gamma + \ln \alpha )\, ,
\end{equation}
  and
\begin{equation}
\int_0^{\infty }  \exp(- \alpha \, x^2)\, \ln x \,  dx = -  \left({\frac{\pi }{16\alpha }}\right)^{1/2}  [ \gamma + \ln (4\alpha )]
\end{equation}
where  $\gamma $ is the Euler-Mascheroni constant,  $\alpha >0$.

\subsection{ Thermodynamical  extremum principles in  Smoluchowski processes}

Now we can pass to the point (***) and discuss  the role of the Helmholtz extremum principle which  often takes the role of more familiar
min/max entropy principle in random motion, c.f. \cite{gar,mackey,gar2}.
 Given a  probability density $\rho(x,t)$  solving the  Fokker-Planck equation
\begin{equation}
\partial _t\rho = D\triangle \rho -  \nabla \cdot ( b \rho ) \, .
\end{equation}
We introduce  $u = D\ln \rho $ and $v=b - u$ which obeys $\partial _t \rho = - \nabla (\rho v)$.\\
The  Shannon  entropy   of $\rho $
\begin{equation}
{\cal{S}}(t)  = -\langle \ln \rho \rangle
\end{equation}
  typically is not a conserved quantity. We impose  boundary restrictions
that $\rho, v\rho, b\rho $ vanish  at spatial infinities or  other  integration interval borders.
We consider:
\begin{equation}
  D \dot{\cal{S}}  =  \left< {v}^2\right>
    -  \left\langle {b}\cdot {v}
 \right\rangle  \label{balance}  \, .
\end{equation}
We may   pass to time-independent drift fields
  and  set  $ b = \frac{f}{m\beta }$,  $j \doteq v\rho $,  $ f = - \nabla V $  plus   $D=k_BT/m\beta $. Then:
\begin{equation}
\dot{\cal{S}} = \dot{\cal{S}}_{int} + \dot{\cal{S}}_{ext}
\end{equation}
where
\begin{equation}
k_BT \dot{\cal{S}}_{int}  \doteq m\beta   \left<{v}^2\right> \geq 0
\end{equation}
stands for the {  entropy production} rate, while
\begin{equation}
k_BT \dot{\cal{S}}_{ext} =   \dot{\cal{Q}} =  -  \int {f} \cdot {j}\,  dx =
- m\beta  \left\langle {b}\cdot {v}
 \right\rangle
\end{equation}
 (as long as negative  which is not a must)  may be  interpreted as the  {  heat dissipation rate}:
 that in view of  $\dot{{\cal{Q}}}= - \int {f}\cdot {j}\,  dx$.

 Notice that because of  $T\dot{S}\doteq k_BT \dot{\cal{S}}$  we do have
\begin{equation}
 T\dot{S}_{int} = T\dot{S} - \dot{Q} \geq 0 \Rightarrow T\dot{S} \geq \dot{Q}\, .
\end{equation}
In view  of $j = \rho v = {\frac{\rho }{m\beta }} [ f - k_BT \nabla \ln \rho ] \doteq  - {\frac{\rho }{m\beta }}\nabla \Psi
$  i.e.  $v= - (1/m\beta ) \nabla \Psi $ and  $f=-\nabla V$, we can  introduce\\
\begin{equation}
\Psi = V + k_BT \ln \rho
\end{equation}
whose mean value stands for   the {   Helmholtz free  energy} of   the random  motion
\begin{equation}
F \doteq \left< \Psi \right> = U - T S \, .
\end{equation}
Here  $S \doteq k_B {\cal{S}}$ and an internal energy is $ U = \left< V\right>$.
Since we assume  $\rho $  and
$\rho V v$ to  vanish at the integration volume boundaries,  we get\\
\begin{equation}
\dot{F}  =   \dot{Q} - T\dot{S} =  - (m\beta )
 \left<{v}^2\right> = - k_BT \dot{\cal{S}}_{int} \leq 0 \, . \label{helm}
\end{equation}
Clearly, $F$ decreases as a function of time  towards its  { minimum},
or  remains constant.

Let us consider the stationary regime   $\dot{\cal{S}} =0$ associated with an  invariant density $\rho _{*}$.
 Then,    $$b=u = D \nabla  \ln \rho _{*} $$ and
\begin{equation}
 -(1/k_BT)\nabla V = \nabla \ln\, \rho _{*}  \Longrightarrow \rho _{*} = {\frac{1}Z} \exp[ - V/k_BT]\, .
 \end{equation}
Hence
\begin{equation}
\Psi _{*} = V + k_BT \ln \rho _{*}  \Longrightarrow \langle \Psi _{*} \rangle =
 - k_BT \ln Z  \doteq  F_{*}
 \end{equation}
   with   $Z= \int \exp(-V/k_BT) dx$.
 $F_*$ stands for   a  minimum  of  the Helmholtz
free  energy $F$. Because of
\begin{equation}
Z= \exp (-F_*/k_BT)
\end{equation}
 we have
\begin{equation}
\rho _* =
\exp[(F_* - V)/k_BT]
\end{equation}

Therefore,   the {  conditional  Kullback-Leibler   entropy},
of the density $\rho $    relative to an equilibrium density $\rho _*
$ acquires the form
\begin{equation}
 k_BT {\cal{H}}_c  \doteq  - k_BT \int \rho \ln
({\frac{\rho }{\rho _*}})dx = F_* - F \, .
\end{equation}

In view of the concavity property of
the function $f(w) = - w\ln w$,  ${\cal{H}}_c$ takes only
negative values, with  a maximum at $0$. We have   $F_*\leq F$ and
 $k_BT \dot{\cal{H}}_c = - \dot{F}  \geq 0$.  ${\cal{H}}_c$  is bound to  grow monotonically
 towards $0$,  while  $F$ drops down to   its
 minimum  $F_*$  which is reached  for $\rho _*$ of Eqs.~(76) and (77).

\subsection{Shannon entropy in quantum systems}

Presently, we pass to the point (****)  raised before in  section $3.3$.
For   probability distributions $p(x)$ on $R$, with any finite mean value,  whose   variance is  fixed
at a the prescribed  value  $\sigma ^2$, we  have
$ S(p)\leq  {\frac{1}2}  \ln (2\pi e \sigma ^2)$.
$S(p)$  becomes maximized if and only if $p$ is a Gaussian with  that variance.

Given  an $L^2(R)$-normalized function  $\psi (x)$, \,
$({\cal{F}}\psi )(p)$ is  its Fourier transform. The corresponding
probability densities:
 $\rho (x) = |\psi (x)|^2$ and $\tilde{\rho }(p) = |({\cal{F}}\psi )(p)|^2$
give rise to  position and momentum  information
(differential, e.g. Shannon) entropies:
\begin{equation}
{\cal{S}}(\rho )\doteq S_q =  - \langle \ln \rho \rangle = - \int
\rho (x) \ln \rho (x)  dx
\end{equation}
$$
    {\cal{S}}(\tilde{\rho })\doteq
S_p= - \langle  \ln \tilde{\rho }\rangle =   - \int \tilde{\rho
}(p) \ln \tilde{\rho }(p)  dp
$$

  For the sake of clarity, we
 use dimensionless quantities, although there exists a consistent procedure for handling
  dimensional quantities in the Shannon entropy definition, c.f. Eq.(8), \cite{gar,gar2}.
We assume both entropies to take finite values to yield an    entropic uncertainty relation, \cite{stam,mycielski}:
\begin{equation}
S_q + S_p \geq  (1 + \ln \pi ) \, .  \label{uncertainty}
\end{equation}

If  we define the squared  standard
deviation   value  for an observable $A$ in a pure state $\psi $
as $(\Delta A)^2 = (\psi , [A - \langle A\rangle ]^2 \psi )$ with
$\langle A \rangle  = (\psi , A\psi)$, then for the  position $X$
and momentum $P$ operators we have ($\hbar \equiv 1$):
\begin{equation}
\Delta X \cdot \Delta P \geq  {\frac{1}{2\pi e}} \,
 \exp[{\cal{S}}(\rho )  + {\cal{S}}(\tilde{\rho })] \geq {\frac{1}{2}} \,  \label{un}
 \end{equation}
 After the Fourier transformation,  taking into   account
 the entropic uncertainty relation  we have ($\sigma ^2$  stands for the variance)
\begin{equation}
4 \tilde{\sigma }^2 \geq  2(e\pi )^{-1}  \exp[-2  \langle \ln
\tilde{\rho }\rangle ] \geq  (2e\pi ) \exp[ 2 \langle \ln  \rho
\rangle ] \geq \sigma ^{-2} \label{chain1}
\end{equation}

For the  momentum operator $P$ that is conjugate  to the
position operator  $X$ in the adopted dimensional convention
$\hbar \equiv 1$. Setting $P= - i d/dx$ and presuming that all
averages are finite,  we get:
$[\langle P^2\rangle - \langle P\rangle ^2] =  (\Delta P)^2=
\tilde{\sigma }^2 $.
The  standard indeterminacy relationship
 $\sigma  \cdot \tilde{\sigma }\geq (1/2)$  follows.  In the stationary  state $\langle P\rangle =0$.

There  is  a subtlety to be mentioned, if we wish to extend the above reasoning to Calogero-type models, \cite{karw}.
We have  \it effectively \rm  considered them  in $L^2(R^+)$ instead of $L^2(R)$, to  establish links with RMT-spacing distributions.
Such restriction of the degenerate quantum model to its non-degenerate projection on $R^+$ is not an  innocent  step  in the
fully-fledged quantum formalism. For example,  there is no  standard  momentum observable (self-adjoint operator of the form
 $ -i\hbar \partial _x$) on
$L^2(R^+)$  alone.

A discussion of this issue can be found in Ref. \cite{karw}.  We  point out that the degeneracy-induced  $R= R^-\bigoplus R^+$    decomposition
 of $L^2(R)$ into $L^2(R^-)\bigoplus L^2(R^+)$,  for the Calogero quantum system,  makes legitimate the usage of the standard momentum  observable.
 To this end, the sufficient and necessary condition  is that  the Hamiltonian  $H$  decomposes as  well  $H= H_{-}\bigoplus H_+$ where  $H_-$ and $H_+$
 are self-adjoint on  their  domains in $L^2(R^-)$  and $L^2(R^+)$ respectively. We have here fulfilled the conditions for a permanent dynamical
 confinement on $L^2(R^+$, as secured by $H_-$,  set in coexistence with the standard $L^2(R)$ momentum observable.

 Another possible suggestion is to accept that a momentum-type operator needs not to be self-adjoint but merely symmetric. The uncertainty relations
are known  to hold true in this case (by the way there are many mathematical references to closely related issue of the time-frequency indeterminacy).

The  Shannon entropy in the  position space, and the so-called Leipnik entropy which coincides with the sum $S_q + S_p$, were investigated
numerically for  excited states of   various quantum  systems, \cite{yanez}-\cite{massen}.
 The typical  observation is that the  entropy values,   evaluated for the ground state probability densities, are minimal.
An explicit check has never  been  made for the Calogero system, but we expect this generic
pattern of behavior to be respected.

\end{document}